\documentclass[letterpaper, 11pt, conference]{styles/ieeeconf}

\IEEEoverridecommandlockouts

\usepackage{cite}
\usepackage{amsmath,amssymb,amsfonts}
\usepackage{graphicx}
\usepackage{xcolor}
\usepackage{booktabs}
\usepackage{url}
\usepackage{tikz}
\usetikzlibrary{positioning,arrows.meta,fit,calc}

\newcommand{\toolName}{\texttt{pytest-gpu-proof}}

\title{\LARGE \bf \toolName: Enabling Cloud-CPU Continuous Integration for GPU Code with Local GPU Attestation}

\author{Brian Plancher$^{1}$%
\thanks{$^{1}$Brian Plancher is with Dartmouth College, Hanover, NH, USA. (Email:
{\tt\small plancher@dartmouth.edu})}%
\thanks{This project was supported by the National Science Foundation (Award 2411369) and the Toyota Research Institute. Any opinions, findings, conclusions, or recommendations expressed in this material are those of the author and do not necessarily reflect those of the funding organizations.}
}

\begin{document}

\maketitle
\thispagestyle{empty}
\pagestyle{empty}

\begin{abstract}
GPU acceleration is now routine across robotics, but cloud-hosted GPU continuous integration (CI) runners are expensive, resulting in severe under-testing of GPU-accelerated code. We present \toolName{}, an open-source \texttt{pytest} plugin offering a practical middle ground. Tests can be run on a local machine, signed with a receipt of exactly what ran and what it produced, and integrated into standard CPU CI workflows (e.g., \texttt{GitHub} Actions). The tool is open source and on PyPI, and we are actively integrating it across our lab's software stack.
\end{abstract}

\section{Introduction}
GPU acceleration is now routine across robotics and the trend is only accelerating. In fact, papers with ``GPU'' in the title or abstract grew from 25 at IROS and ICRA 2023 combined to 89 at IROS 2025 and ICRA 2026 combined. This includes work developed by industry and academia on topics from across the robotics stack, including: simulation, computer vision, estimation, mapping, planning, control, end-to-end learning and more~\cite{xplore,papercept-icra-26}. This is all occurring at a time when the number of paper submissions across academia is rapidly rising and AI coding tools have not only supported this trend but also led to a proliferation of poorly tested code that is hard to maintain~\cite{billard2026surviving,wang2026maintaincoder,ghammam2026ai,lange2025sakana}.

This underscores that academia, and software engineering more broadly, needs tools that can help low-resourced developers (e.g., common in small academic labs) maintain and test their code through industry best practices like continuous integration (CI)~\cite{duvall2007continuous,shahin2017continuous}.

\begin{figure}[!t]
\centering
\begin{tikzpicture}[
  font=\scriptsize,
  node distance=0.28cm and 0.5cm,
  box/.style={draw, rounded corners=1pt, inner sep=3pt, align=left,
              minimum width=3.35cm},
  gpu/.style={box, fill=orange!10},
  ci/.style={box, fill=blue!7},
  flow/.style={-{Stealth}, thick},
]
\node[gpu] (run)
  {\textbf{local GPU run} (pytest, one-time)\\
   over source tree + marked GPU tests};
\node[gpu, below=of run] (rcpt)
  {canonical JSON receipt:\\
   fingerprint $\cdot$ commit $\cdot$ outcomes $\cdot$ skips $\cdot$ env};
\node[gpu, below=of rcpt] (sign)
  {\textbf{Ed25519 sign} (developer's SSH key)};
\node[box, fill=violet!8, below=of sign] (art)
  {committed artifact \ \texttt{gpu-proof.json}};
\node[ci, below=of art] (verify)
  {\textbf{CPU-only CI, every push:} verify\\
   signature (signer's published GitHub keys)\\
   $\cdot$ source fingerprint at HEAD\\
   $\cdot$ commit ancestry $\cdot$ exact outcome + skip set\\
   $\cdot$ freshness $\cdot$ shard-carry policy};
\foreach \a/\b in {run/rcpt, rcpt/sign, sign/art, art/verify}
  \draw[flow] (\a) -- (\b);
\node[draw=red!60, dashed, thick, inner sep=3pt,
      fit=(run)(rcpt)(sign)] (boundary) {};
\node[anchor=south west, font=\tiny, text=red!60] at (boundary.north west)
  {trust boundary: everything here is self-reported by the signer};
\end{tikzpicture}
\caption{The (possibly expensive or time consuming) GPU run is local and is signed. Every following (cloud-based) CI run verifies the signed attestation in seconds and fails on any drift between the attested run and the current tree.}
\vspace{-15pt}
\label{fig:receipts}
\end{figure}
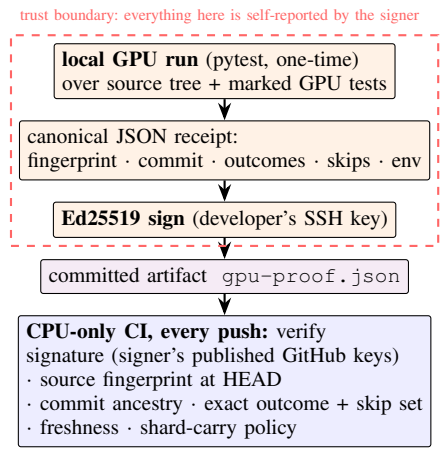

At the same time, while popular cloud-based code repositories like \texttt{GitHub} support CI workflows~\cite{duvall2007continuous,shahin2017continuous} that can run popular testing suites (e.g., \texttt{pytest}~\cite{pytest}), running GPU programs in such environments often requires the use of costly advanced features~\cite{githubCI}. For all but the largest corporations, cloud-based GPU CI is thus generally avoided. This means that open-source, GPU-accelerated robotics academic software is often severely under-tested despite the fact that this code often contains difficult to detect errors (e.g., data-dependent race conditions~\cite{yu2005racetrack,betts2012gpuverify}).

We address this issue with \toolName{}, a plugin for \texttt{pytest} that enables GPU tests to be run locally, cryptographically signed to provide attestation of exactly what ran and what it produced, and integrated into CPU-based CI workflows on popular repositories like \texttt{GitHub}. Our plugin is lightweight and easy to install as it only requires \texttt{pytest} and an SSH key, and its own CI runs at 100\% test coverage. We hope this tool helps enable the explosion of GPU-based robotics software to be better maintained and more reliable. We have released our tool open source under an MIT license at
\url{https://github.com/A2R-Lab/pytest-gpu-proof}, with version 0.4.0
on PyPI (\texttt{pip install pytest-gpu-proof}).

\section{Plugin Design and Operation}
\toolName{} operation can be summed up in one clause: run locally, sign, and verify everywhere. 

As shown in Fig.~\ref{fig:receipts}, tests that need a GPU are selected with a pytest marker and run locally. We then record the collected test set, per-test outcomes and durations, expected skips, git commit and tree state, the environment and reported GPU, and a SHA-256 fingerprint over the configured source scope (by default every tracked file). This record is serialized as a JSON file and signed with the developer's existing Ed25519 SSH key (verified later against published \texttt{GitHub} keys). 
The receipt is then committed with the code, and the CI (local or cloud) performs one quick, CPU-only step to verify the receipt. That is, verification recomputes the source fingerprint at HEAD, requires the recorded commit to be the current commit or an ancestor, requires every recorded test to have passed, compares the \emph{exact} expected-skip set, and applies freshness and clean-tree policies to ensure no edits were made after the signed test receipt.

Beyond simple GPU test markers, our plugin includes a comparison feature that pairs a GPU kernel under test with a (user provided) CPU or GPU reference implementation and records the outcome in the receipt.

\toolName{} supports a variety of different repository policies through the use of a small declarative policy file that e.g., pins the required fingerprint scope, maximum
receipt age, recording mode, and an optional test manifest. To ensure that a misspelling cannot silently weaken the policy, unknown policy fields fail verification. For large suites, receipts can be recorded in \emph{shards} on different machines and merged, with a carry-forward policy that lets shards whose fingerprinted inputs are unchanged ride along without rerunning. Policies can also differentiate between acceptable CI for a single commit as compared to a release  (e.g., requiring a full test set run before a release). Finally, policies can specify whether the signer policy is \emph{open}, which enables a valid receipt from any GitHub user (e.g., to support contributor-signed pull requests), or \emph{restricted}, which allowlists specific users for added security (e.g., around releases).

\section{What a receipt does and does not prove}

With \toolName{}, the claim ``the GPU tests pass'' is pinned to an exact source tree, commit, test list, outcome set, and date, and is re-checked on every push, so accidental staleness (the dominant failure mode in practice) is caught immediately. However, a receipt remains only an \emph{attestation by a keyholder}, not cryptographic proof of GPU execution as a dishonest signer can sign anything. That being said, this is the same trust model as a signed git tag or release artifact~\cite{newman2022sigstore}, and a \emph{restricted} signing policy limits that trust to a chosen set of developers.

\section{Conclusion and Future Work}

\toolName{} enables CI of GPU projects while keeping expensive GPU compute on local hardware. We hope the workshop discussion will help refine the policy vocabulary and we welcome adversarial feedback on our plugin design and trust model.

\section*{Acknowledgments}
We used LLM tools, including Codex and Claude Code, to assist with software implementation, proofreading, and review. All final text and references were edited and reviewed by humans.

\bibliographystyle{styles/IEEEtran_new}
\bibliography{refs}

\end{document}